\documentclass[aps,prc,twocolumn,showpacs,superscriptaddress]{revtex4-2}

\usepackage{dcolumn}   
\usepackage{bm}        

\usepackage{graphicx}  
\usepackage[justification=justified,format=plain]{caption}
\usepackage[justification=justified,format=plain]{subcaption}
\usepackage{amssymb}   

\usepackage{lineno}
\usepackage{multirow}

\usepackage{xcolor,cancel}

\begin{document}



\title{Structure studies of $^{257}$Db through combined $\alpha$, $\gamma$ and internal-conversion-electron spectroscopy}

\author{P. Brionnet}
\email{corresponding author, present affiliation RIKEN, Nishina Center, Wako, Japan}
\affiliation{Universit\'e de Strasbourg, CNRS, IPHC UMR 7178, F-67000 Strasbourg}
\author{A. Lopez-Martens}
\affiliation{IJClab, Universit\'e Paris Saclay and CNRS, F-91405 Orsay, France}
\author{O. Dorvaux}
\affiliation{Universit\'e de Strasbourg, CNRS, IPHC UMR 7178, F-67000 Strasbourg}
\author{Z. Asfari}
\affiliation{Universit\'e de Strasbourg, CNRS, IPHC UMR 7178, F-67000 Strasbourg}
\author{M.L. Chelnokov}
\affiliation{FLNR, JINR, Dubna, Russia}
\author{V.I. Chepigin}
\affiliation{FLNR, JINR, Dubna, Russia}
\author{H. Faure}
\affiliation{Universit\'e de Strasbourg, CNRS, IPHC UMR 7178, F-67000 Strasbourg}
\author{B. Gall}
\affiliation{Universit\'e de Strasbourg, CNRS, IPHC UMR 7178, F-67000 Strasbourg}

\author{K. Hauschild}
\affiliation{IJClab, Universit\'e Paris Saclay and CNRS, F-91405 Orsay, France}

\author{A.V. Isaev}
\affiliation{FLNR, JINR, Dubna, Russia}
\affiliation{Dubna State University, Dubna, Russia}
\author{I.N. Izosimov}
\affiliation{FLNR, JINR, Dubna, Russia}
\author{A.A. Kuznetsova}
\affiliation{FLNR, JINR, Dubna, Russia}
\author{O.N. Malyshev}
\affiliation{FLNR, JINR, Dubna, Russia}
\affiliation{Dubna State University, Dubna, Russia}
\author{R.S Mukhin}
\affiliation{FLNR, JINR, Dubna, Russia}
\author{J. Piot}
\affiliation{GANIL, Cean, France}
\author{A.G. Popeko}
\affiliation{FLNR, JINR, Dubna, Russia}
\affiliation{Dubna State University, Dubna, Russia}

\author{Yu.A. Popov}
\affiliation{FLNR, JINR, Dubna, Russia}
\author{K. Rezynkina}
\affiliation{IJClab, Universit\'e Paris Saclay and CNRS, F-91405 Orsay, France}
\author{B. Sailaubekov}
\affiliation{FLNR, JINR, Dubna, Russia}
\affiliation{L.N. Gumilyov Eurasian National University, Astana, Kazakhstan}
\author{E.A. Sokol}
\affiliation{FLNR, JINR, Dubna, Russia}
\author{ A.I. Svirikhin}
\affiliation{FLNR, JINR, Dubna, Russia}
\affiliation{Dubna State University, Dubna, Russia}

\author{M.S. Tezekbayeva}
\affiliation{FLNR, JINR, Dubna, Russia}
\affiliation{The Institute of Nuclear Physics, 050032, Almaty, The Republic of Kazakhstan}
\affiliation{Dubna State University, Dubna, Russia}
\author{A.V. Yeremin \textsuperscript{\textdagger}} 
\affiliation{FLNR, JINR, Dubna, Russia}
\affiliation{Dubna State University, Dubna, Russia}
\author{N.I. Zamyatin} 
\affiliation{VBLHEP, JINR, Dubna, Russia}


\begin{abstract}

This work reports on the study of the decay properties along the $^{257}$Db decay chain using the GABRIELA setup. The first observation of a high-K isomer in $^{257}$Db is presented. In addition, an unreported $\alpha$-decay branch in $^{249}$Md has been evidenced, allowing to constrain the differences in energy of the $\alpha$-decaying levels in $^{249}$Md, $^{253}$Lr and $^{257}$Db. Finally, the combination of the observed fine structure $\alpha$-decay from the high-spin state in $^{257}$Db with the first observation the internal decay in $^{253}$Lr requires a revision of level and decay scheme. In particular, a change of parity for the high-spin state from 9/2$^{+}$ to 9/2$^{-}$ in the $^{257}$Db is suggested, and the implications of such a change are also discussed.

\end{abstract}






\maketitle




\section{Introduction}
\label{Intro}

The limits of nuclear existence remain one of the key challenges in modern nuclear physics, with ongoing studies in several laboratories worldwide aimed at exploring new elements and isotopes~\cite{SRILAC,SHE_factory,SHANS2,gates2024towards}. However, these studies are hindered by limited statistics due to the extremely low production cross sections.
In contrast, lighter nuclei ($100 \leq Z \leq 114$) exhibit significantly higher production rates, allowing for more detailed investigations of their structure. These studies provide valuable insights into the organisation of nuclear matter under extreme charge and mass conditions.

Deformed superheavy nuclei (SHN) around neutron numbers $N = 152$ and $162$ are particularly interesting, as some of the proton and/or neutron orbitals they occupy are predicted to originate from those responsible for the stabilisation of SHN in the so-called ``Island of Stability''. The presence of both low- and high-spin orbitals at the Fermi surface of these deformed nuclei leads to the widespread occurrence of spin and high-$K$ isomers. The decay of these isomers provides a distinct experimental signal, with properties that serve as fingerprints of the underlying microscopic structure.

\begin{figure}[ht!]
      \centering
      \includegraphics[width=0.81\linewidth]{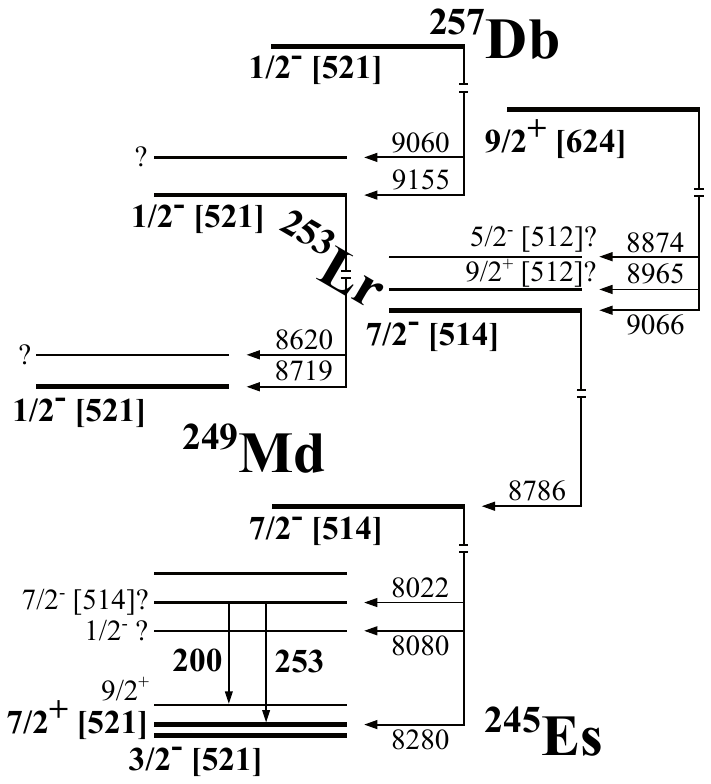}
      \caption{Partial decay scheme of the $^{257}$Db decay chain obtained from previous experiments~\cite{hessberger2001decay,gates2008comparison,streicher2006synthesis,pbrionnet}}
      \label{Level}
\end{figure}

In this paper, we report a new study of the decay properties of $^{257}$Db using the GABRIELA multi-detector array~\cite{GABRIELA}, coupled with the SHELS separator~\cite{SHELS}. Two main results are presented: the first observation of a high-$K$ isomer in $^{257}$Db and the connection between the parallel $\alpha$-decay chains linking the ground state and low-spin isomer in $^{257}$Db, $^{253}$Lr, and $^{249}$Md, as shown in Fig.~\ref{Level}, through a previously unreported $\alpha$-decay branch from the 1/2$^-$ state in $^{249}$Md.

The peculiar decay properties of the high-spin ground state of $^{257}$Db have also been confronted with Geant4 Monte Carlo simulations, leading to a new spin and parity assignment. The implications for the sequence of proton states above $Z$~=~100 and the relative position of high-$j$ shells are discussed. 

\section{Experiment}
\label{Expt}

$^{257}$Db nuclei were produced in three separate experimental runs via the fusion-evaporation reaction $^{50}$Ti($^{209}$Bi,$x$n)$^{259-x}$Db, using the two-neutron evaporation channel.

The $^{50}$Ti beam was generated by the ECR-4M ion source~\cite{ECR} using the Metal Ions from Volatile Compounds (MIVOC) method~\cite{MIVOC} and subsequently injected into the U400 Cyclotron~\cite{U-400} at the Flerov Laboratory of Nuclear Reactions (FLNR), Dubna.

The upgraded GABRIELA setup~\cite{chakma2020gamma} was employed to detect evaporation residues and their subsequent decays following selection and transportation through the SHELS separator~\cite{SHELS}. Incoming ions were identified at the focal plane using two Time-Of-Flight (TOF) detectors, comprising a single emissive foil and two micro-channel plate detectors for secondary electron emission detection.

The focal plane detection system consists of two detector arrays. First, a silicon detector box is used for detecting incoming nuclei and their subsequent decays. The ions are implanted into a 10$\times$10 cm$^{2}$ Double-Sided Silicon Strip Detector (DSSD) with 128 strips on each side, providing 16,384 pixels. Particles escaping the DSSD are captured by eight 6$\times$5 cm$^{2}$ DSSDs, each featuring 16 strips per side, arranged upstream from the implantation detector in a tunnel configuration.

The silicon box is surrounded by five High-Purity Germanium (HPGe) detectors for $\gamma$-ray detection. A Clover HPGe detector is positioned directly behind the implantation detector, while four coaxial Ge detectors are arranged in a cross configuration around the tunnel detectors to ensure the highest detection efficiency possile~\cite{chakma2020gamma}. All five detectors are equipped with BGO shields to suppress Compton background.

Calibration of the GABRIELA detectors was performed using $^{164}$Dy and $^{170}$Er targets, producing $^{209-210}$Ra and $^{216-217}$Th isotopes, along with standard radioactive sources. The measured Full Width at Half Maximum (FWHM) of the 7.92 MeV $\alpha$-decay of $^{216}$Th was 31~keV. The DSSD detector thresholds ranged from $\sim$100 to 300 keV, depending on the amplification dynamic range of the signals detected on the back-side strips~\cite{lopez2019}.

The $^{209}$Bi targets used in the three production runs of $^{257}$Db were grouped into six sets of metallic bismuth targets, each deposited on a 3~$\mu$m Al backing. The measured average density of each set ranged from 380 to 510~$\pm$20 $\mu$g$\cdot$cm$^{-2}$. A beam intensity of 300-350 pnA was used, with the mid-target energy optimised to match the previously measured maximum of the two-neutron evaporation channel~\cite{gates2008comparison}. The extracted mid-target energies ranged from 237.7 to 238.8 MeV, depending on the target thickness. Measurements of the production cross sections and their comparison to previous studies have been reported by A. Lopez-Martens \textit{et al.}~\cite{lopez2019}.

\section{Results}
\label{RESULTS}

The identification of the nuclei produced in the reaction was performed using time and energy correlation of the successive $\alpha$ decays registered in each pixel of the DSSD.

\begin{figure}[ht!]
	\centering
	\includegraphics[width=0.97\linewidth]{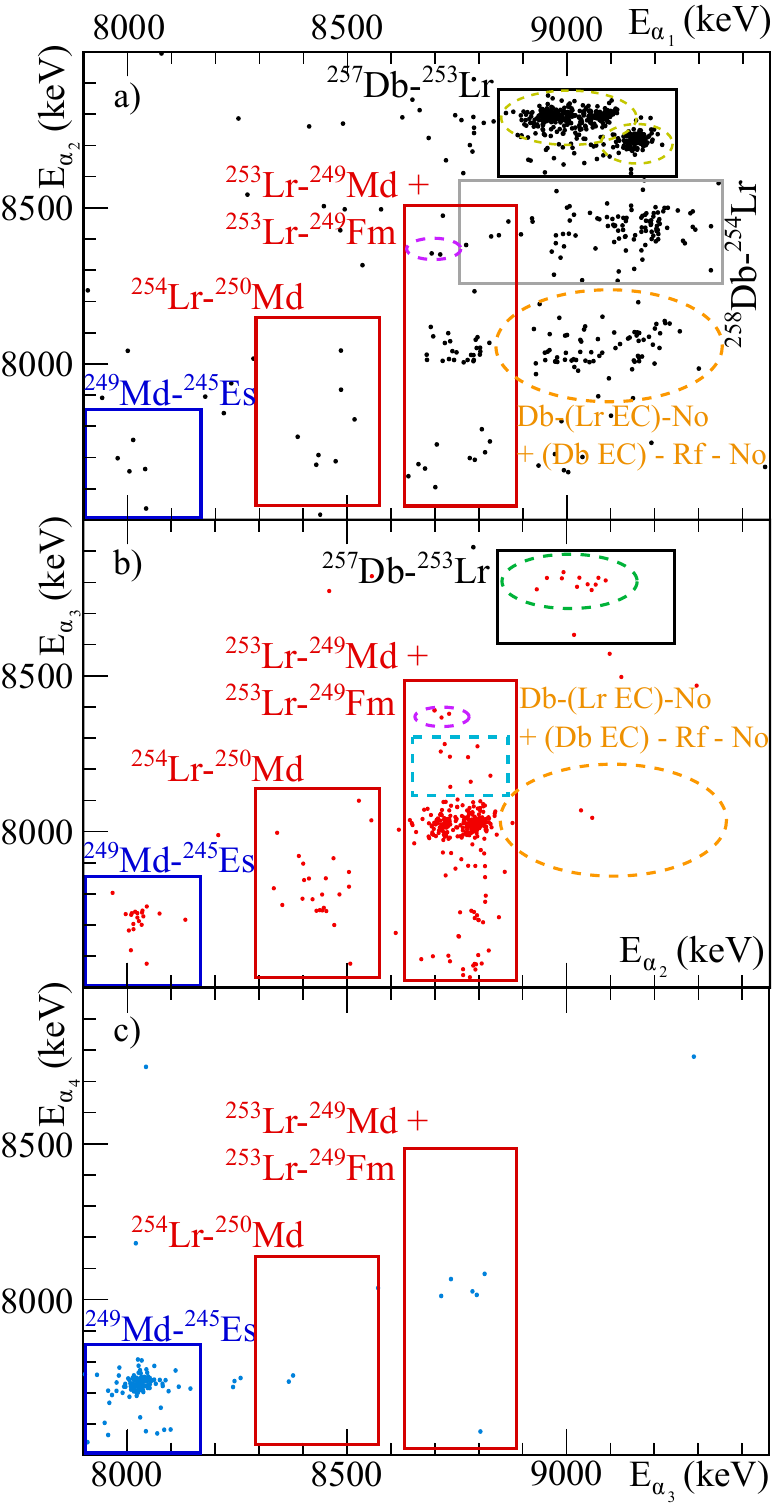}
    \caption{Genetic correlations between successive decays in the DSSD: correlation in energy between (a) the first and second generations, (b) the second and third generations, and (c) the third and fourth generations. The correlation time gate applied is $\Delta$t$<$900~s between each generation. The different boxes and dashed circles correspond to the identification and properties observed in this work (see text for more details).}
    \label{Eam_Ead}
\end{figure}

\begin{figure*}[!t]
    \centering
    \includegraphics[width=0.95\linewidth]{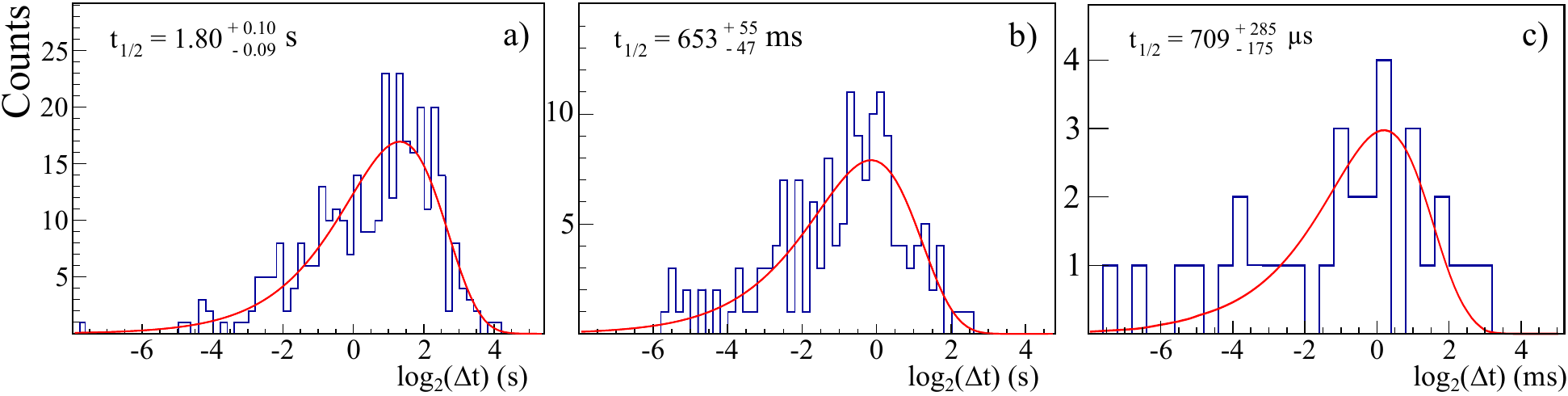}
    \caption{Decay time in $\log_{2}$ of seconds for the two $\alpha$-decaying states of $^{257}$Db: (a) high-spin state decay time, (b) low-spin state decay time, and in $\log_{2}$ of milliseconds for the newly observed isomeric high-K state in $^{257}$Db: (c).}
    \label{Tlog}
\end{figure*}

The $\alpha$ decays of $^{257}$Db and its daughter $^{253}$Lr are clearly identifiable in Fig.\ref{Eam_Ead}(a) and more faintly visible within the black box in Fig.\ref{Eam_Ead}(b). The high-spin and low-spin decay chain structures are also evident in Fig.\ref{Eam_Ead}(a), highlighted by the two dashed circles within the black box. Additionally, contributions from the one-neutron evaporation channel, characterised by the $\alpha$ decays of $^{258}$Db and $^{254}$Lr (grey box), as well as the electron capture (EC) branches of $^{257,258}$Db and/or $^{253,254}$Lr leading to $^{253,254}$No (orange dashed circles), are observed in Fig.\ref{Eam_Ead}(a-b).
The decays of Lr into Md, corresponding to both neutron evaporation channels, are marked by the red box, while the $\alpha$ decays of $^{249}$Md to $^{245}$Es are located in the blue box.

\subsection{Decay properties of $^{257}$Db}

\begin{figure}[h]
      \centering
     \includegraphics[width=0.9\linewidth]{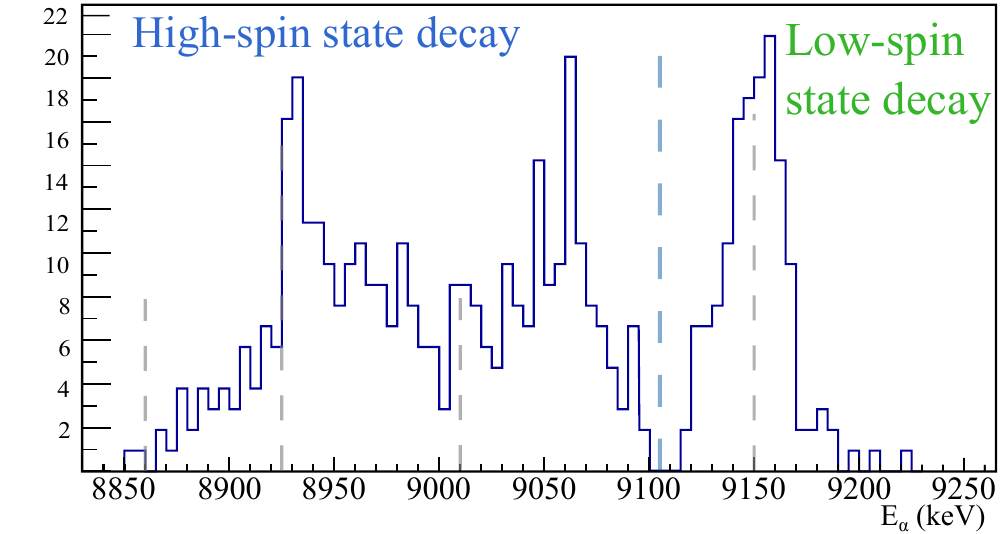}
      \caption{Energy spectrum of the $\alpha$ decays of $^{257}$Db as identified in Fig.~\ref{Eam_Ead} (black box). The gray dashed lines indicate the proposed $\alpha$-decay energies obtained from the Monte Carlo simulations (see the text for details).}
      \label{Ea_fit}
\end{figure}

Earlier studies~\cite{hessberger2001decay,gates2008comparison,streicher2006synthesis} established that the $\alpha$-decay spectrum of $^{257}$Db consists of two distinct groups, characterised by different decay times (see Fig.\ref{Tlog}) and energies (see Fig.\ref{Ea_fit}). The first is a low-energy, double-humped structure spanning 8860--9100~keV, with a half-life of 1.8~s, while the second is a sharp high-energy peak at 9150~keV, with a shorter half-life of 653~ms. The former is associated with the decay of a high-spin state, which undergoes favoured $\alpha$ decays down to $^{245}$Es, whereas the latter is attributed to the decay of the 1/2$^-$[521] Nilsson state, as shown in Fig.~\ref{Level}.

\begin{figure}[h!]
      \centering
     \includegraphics[width=0.95\linewidth]{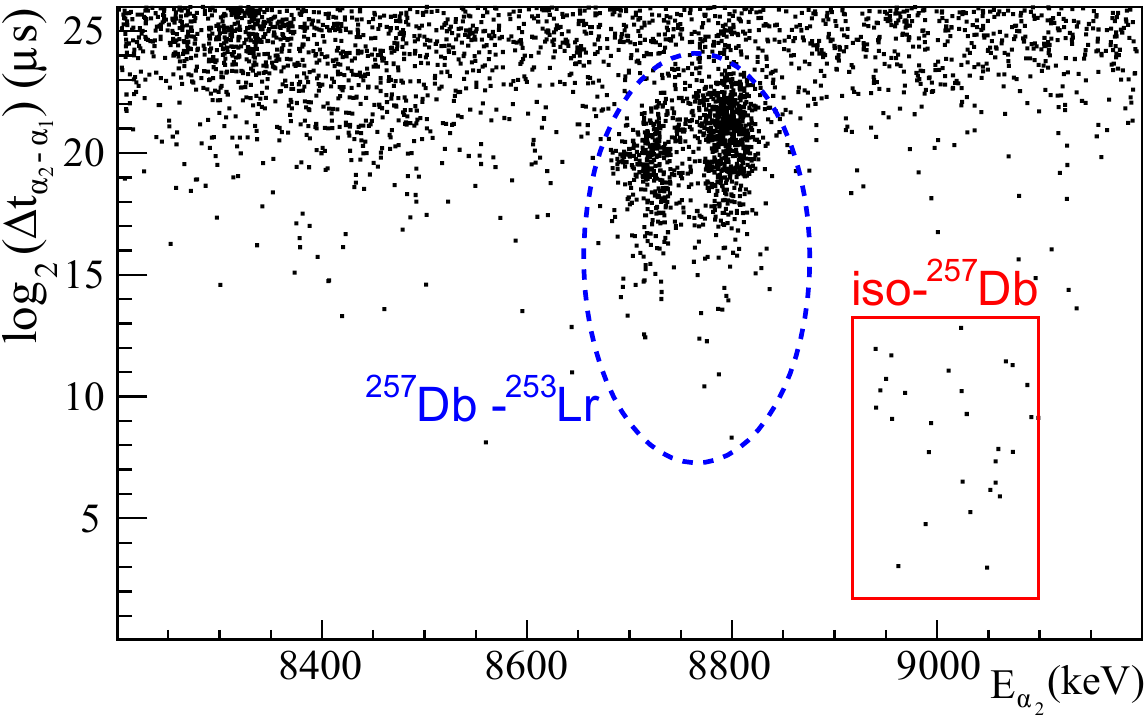}
      \caption{First generation decay time, in $\log_{2}$ of $\mu$s, as a function of the second-generation decay energy. The blue dashed circle corresponds to the standard decay chain of $^{257}$Db, and the red square corresponds to the isomeric state observed above the high-spin state decay in $^{257}$Db.}
      \label{E_iso}
\end{figure}

In addition to the previously reported decays of $^{257}$Db, the decay of a second isomeric state was observed for the first time. The initial indication of these events is visible in Fig.~\ref{Eam_Ead}(b) within the green dashed circle.
These second-to-third generation correlations exhibit characteristics consistent with the previously reported high-spin decays of $^{257}$Db observed in Fig.\ref{Eam_Ead}(a): all events are correlated with $^{257}$Db $\alpha$ decays featuring energies below 9100~keV. These events are also evident in Fig.~\ref{E_iso}, which displays the first-generation decay time as a function of the second-generation decay energy, following any implantation event

\renewcommand{\arraystretch}{1.35}
\begin{table*}[ht!]
      \centering
      \begin{tabular}{ c  c  c  c  c  c  c  c  c }
            & E$_{\alpha}$ (keV) & Feeding (\%) & t$_{1/2}$ & HF & B$_{\alpha}$ (\%) & B$_{SF}$ (\%) & B$_{EC/\beta}$ (\%) & B$_{IT}$ (\%)\\
            \hline
            \multirow{ 4}{*}{$^{257}$Db}& 9009 $\pm$ 31 & 5.0 $\pm$ 1.5 &\multirow{ 3}{*}{1.8 $^{+0.1}_{-0.1}$ s}& 33.9 $^{+3.6}_{-11.5}$ & \multirow{ 3}{*}{97 $\pm$ 4.5} & \multirow{ 3}{*}{3.0 $^{+10}_{-0.5}$} & \multirow{ 3}{*}{-} & \multirow{ 3}{*}{-}\\
            & 8925 $\pm$ 31 & 76.2 $\pm$ 4.5 & & 1.23 $^{+0.08}_{-0.06}$ & & & &\\
            & 8859 $\pm$ 31 & 18.8 $\pm$ 4.5 & & 3.1 $^{+1.0}_{-0.6}$ & & & &\\
            
            & 9145 $\pm$ 31 & - & 0.65 $^{+0.06}_{-0.05}$ s & 1.2 & $\leq$ 100 & - & - & -\\
            \hline
            $^{253}$Lr & 8784 $\pm$ 31 & - & 0.55 $^{+0.03}_{-0.03}$ s & 1.3 & 89 $\pm$ 5.1 & 1.0 $\pm$ 0.7 & 10.0 $\pm$ 4.4 & -\\
            \hline
            $^{253}$Lr & 8713 $\pm$ 31 & - & 2.0 $^{+0.2}_{-0.1}$ s & 2.8 & 83.7 $\pm$ 6.3 & 5.8 $\pm$ 1.7 & 10.5 $\pm$ 4.6 & -\\
            \hline
            $^{249}$Md & 8024 $\pm$ 31& - & 22.3 $^{+1.0}_{-1.0}$ s & 1.5 & 67 $\pm$ 6 & - & 33 $\pm$ 6 & -\\
            \hline
            $^{249}$Md & 8325 $\pm$ 31& - & 1.1 $^{+0.6}_{-0.4}$ s & 28 & 2 $\pm$ 1 & - & - & 98 $\pm$ 1\\
            \hline
      \end{tabular}
      \caption{Decay properties for $^{257}$Db, $^{253}$Lr, and $^{249}$Md isotopes deduced from this work. The $\alpha$-decay energies are extracted from the Monte Carlo simulation performed to account for the summation of converted transitions observed in this work (see the text for details).}
      \label{Recap_decay}
\end{table*}
\renewcommand{\arraystretch}{1}

Two components related to the decay of the $^{257}$Db isotope can be observed in Fig.\ref{E_iso}. First, the event in the blue dashed circle are consistent with the timing and energy measured and listed in Table~\ref{Recap_decay} for the two $\alpha$-decaying state in $^{257}$Db; timing of $^{257}$Db with $^{253}$Lr decay energies. However, the red square corresponds to events prior $\alpha$-decay of $^{257}$Db as the energy of the second generation is consistent with the decays of $^{257}$Db; more specifically only with the high-spin decays of $^{257}$Db for energies lower than 9100~keV. In addition, the decay time of the first generation is much faster than any of the observed decays of $^{257}$Db, and well below any random correlations in Fig.\ref{E_iso}. Therefore, these correlations suggest the existence of decay events preceding the full-energy $\alpha$-decay of the high-spin state of $^{257}$Db. 

\begin{figure}[h!]
      \centering
      \includegraphics[width=1\linewidth]{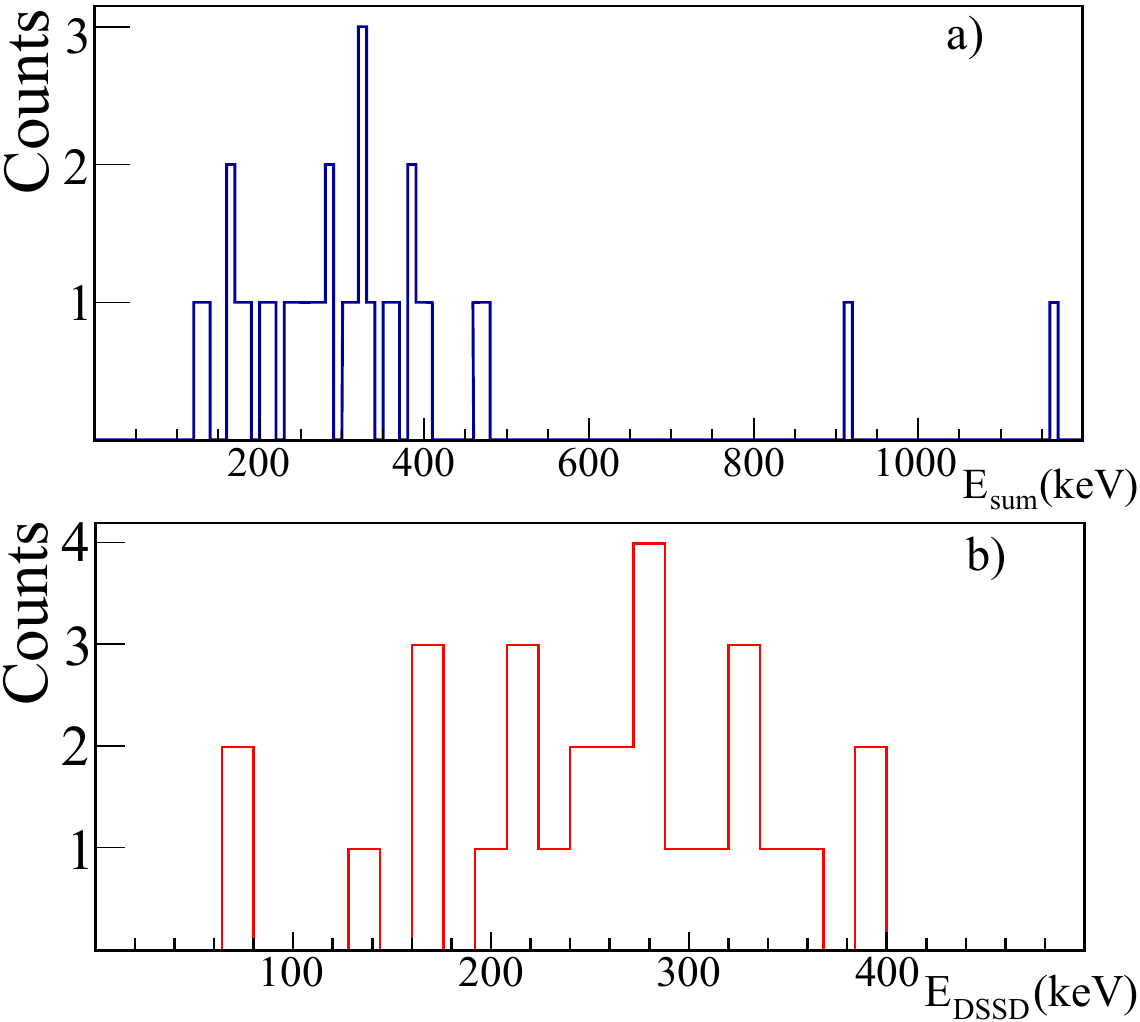}
      \caption{(a) Calorimetric spectrum of isomeric events (see text). (b) Energy spectrum in the DSSD of the isomeric decay as observed in Fig.~\ref{E_iso} (red box).}
      \label{E_calo}
\end{figure}

In total, 33 isomeric decays have been observed, and the half-life of the isomeric state is measured at 0.71~$^{+0.29}_{-0.18}$~ms as seen in Fig.\ref{Tlog}(c). Figure~\ref{E_calo}(b) plots their energy deposition in the DSSD with an average of around 280~keV, and a maximum around 500~keV. The spectrum of the sum of all the energies observed in the isomeric decays (sum of Internal Conversion Electrons (ICEs), auger electrons and X rays in the DSSD, individual ICEs in the tunnel detectors as well as $\gamma$ rays in the Ge detectors) is displayed in Fig.~\ref{E_calo}(a).

The end of point of this calorimetric spectrum provides an estimate of the excitation energy of the isomeric state, which is at least 1.1~MeV.
Given its high excitation energy, the isomeric state is likely a high-K isomer, with a preferred decay towards the high-spin $\alpha$-decaying state of $^{257}$Db, consistent with the observations.

Additionally, one spontaneous fission (SF) event was recorded following the isomeric decay, allowing us to measure an SF branching ratio of 3.0$^{+10}_{-0.5}$~\% for the high-spin state of $^{257}$Db. The decay characteristics of the states identified in this study are summarised in Table~\ref{Recap_decay}.

\begin{figure}[ht!]
      \centering
      \includegraphics[width=1\linewidth]{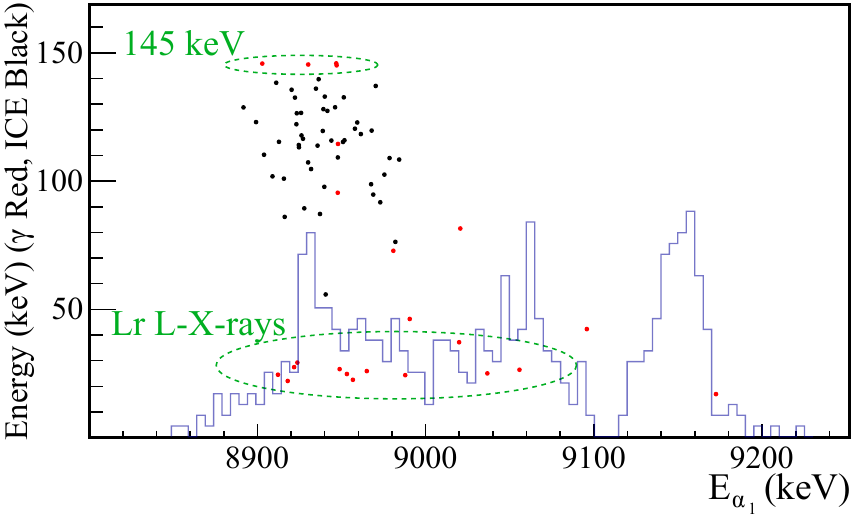}
      \caption{Detected electron and $\gamma$-ray energies as a function of coincident $^{257}$Db $\alpha$-decay energies. The black dots represent the ICEs detected in the tunnel detectors, and the red dots represent the $\gamma$ rays detected in the Ge detectors. The green dashed circles highlight the L-X-rays and the 145~keV transition. The blue spectrum is the spectrum of the first-generation $\alpha$-decay energies from Fig.~\ref{Ea_fit}.}
      \label{Eeg_Ea1}
\end{figure}

Finally, for the first time, internal transitions (IT) have been observed in $^{253}$Lr in coincidence with the $\alpha$ decay of $^{257}$Db ($\Delta t < 3\mu$s). These events are detected exclusively in coincidence with the high-spin $\alpha$ decay of $^{257}$Db, as illustrated in Fig.~\ref{Eeg_Ea1}, which plots the IT event energy as a function of the $^{257}$Db $\alpha$-decay energy.

\begin{figure*}[ht!]
      \centering
     \includegraphics[width=0.95\linewidth]{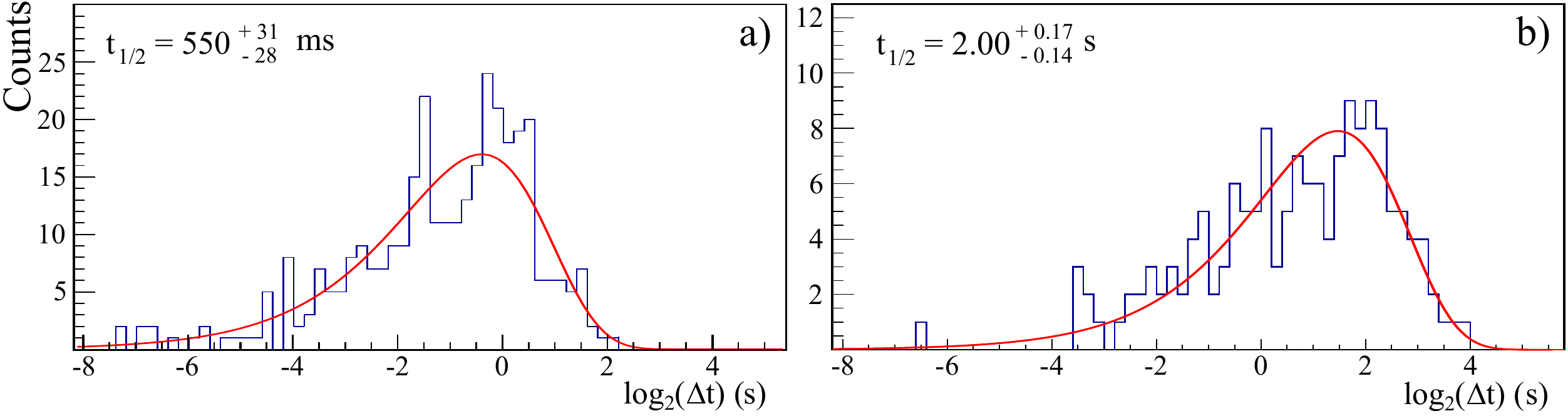}
      \caption{Decay time in $\log_{2}$ of seconds for the two decay state of $^{253}$Lr (a) high-spin state, (b) low-spin state.}
      \label{Tlog_Lr}
\end{figure*}

These transitions are highly converted, as most of the statistics come from events detected in the tunnel detector, with only a few counts in the Germanium detectors, mostly from the X-rays associated with the internal conversion process, as seen in Fig.\ref{Eeg_Ea1}. The 48 events detected in the tunnel detectors are plotted with black dots and range from 80--150~keV. The gamma rays detected are represented by red dots; three events at $\sim$145~keV, as well as L X-rays of Lr, are observed.
The 145~keV transition is only correlated with $E_{\alpha} \sim$8925~keV, which corresponds to the lowest reported $\alpha$-decay energy in Table~\ref{Recap_decay}.
The observed electrons are correlated with $\alpha$ decays below 9000~keV, while the Lr L X-rays are present across the entire energy domain of the high-spin $\alpha$-decaying state. If one assumes that all the electrons above 100~keV correspond to the LMN+ conversion of the 145~keV transition, an internal conversion coefficient of 21$^{+28}_{-11}$ is obtained.

\subsection{Decay properties of $^{253}$Lr}

\begin{figure}[ht!]
      \centering
      \includegraphics[width=0.95\linewidth]{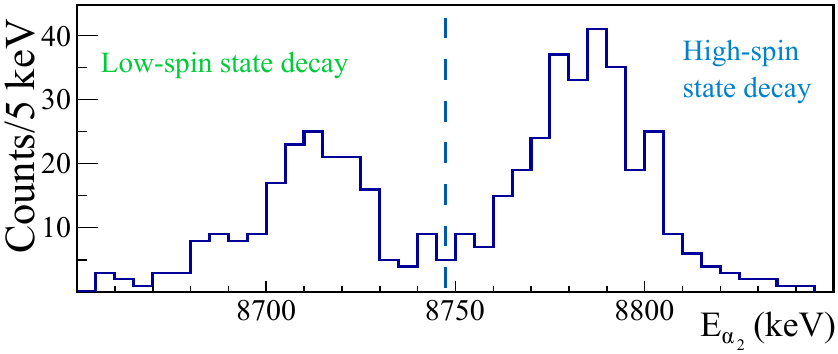}
      \caption{Energy spectrum of the $^{253}$Lr decays as identified in Fig.~\ref{Eam_Ead}(a) (black box). This spectrum is the sum of the high and low-spin state decays of $^{253}$Lr, split around 8748~keV.}
      \label{Ea_Lr}
\end{figure}

The decay energies and times of the two $\alpha$-decaying states of $^{253}$Lr are shown in Fig.~\ref{Tlog_Lr} and \ref{Ea_Lr}.

The measured widths for the two peaks of $^{253}$Lr, shown in Fig.~\ref{Ea_Lr}, as well as the sharp peak of the low-spin state $\alpha$-decay of $^{257}$Db, are comparable to the experimental resolution. These measured widths characteristics, along with the non-observation of $\gamma$ rays or Internal Conversion Electrons (ICEs) in coincidence, suggests that these decays exhibit minimal fine structure.
The half-life for the high-spin state decay of $^{253}$Lr is measured at 0.55(3)~s, while the low-spin state decay half-life is measured at 2.0(1)s. Both values are in good agreement with previously reported results\cite{hessberger2001decay,gates2008comparison,streicher2006synthesis}.

Finally, a measurement of the Electron-Capture (EC) and Spontaneous Fission (SF) branching ratios for $^{253}$Lr can be performed by applying a strict time selection to the first-generation Db $\alpha$ decays, thereby removing contamination from $^{254}$Lr. Indeed, the $\alpha$ decays of $^{257}$Db and $^{258}$Db have overlapping energies, as shown in Fig.\ref{Eam_Ead}(a). Moreover, the EC of $^{254}$Lr results in $^{254}$No, which shares very similar decay characteristics (time and energy) with $^{253}$No, the EC daughter of $^{253}$Lr. By selecting fast Db decay events ($\Delta t \leq$ 750ms), the majority of this contamination can be effectively removed.

\begin{figure}[h!]
	\centering
     \includegraphics[width=0.985\linewidth]{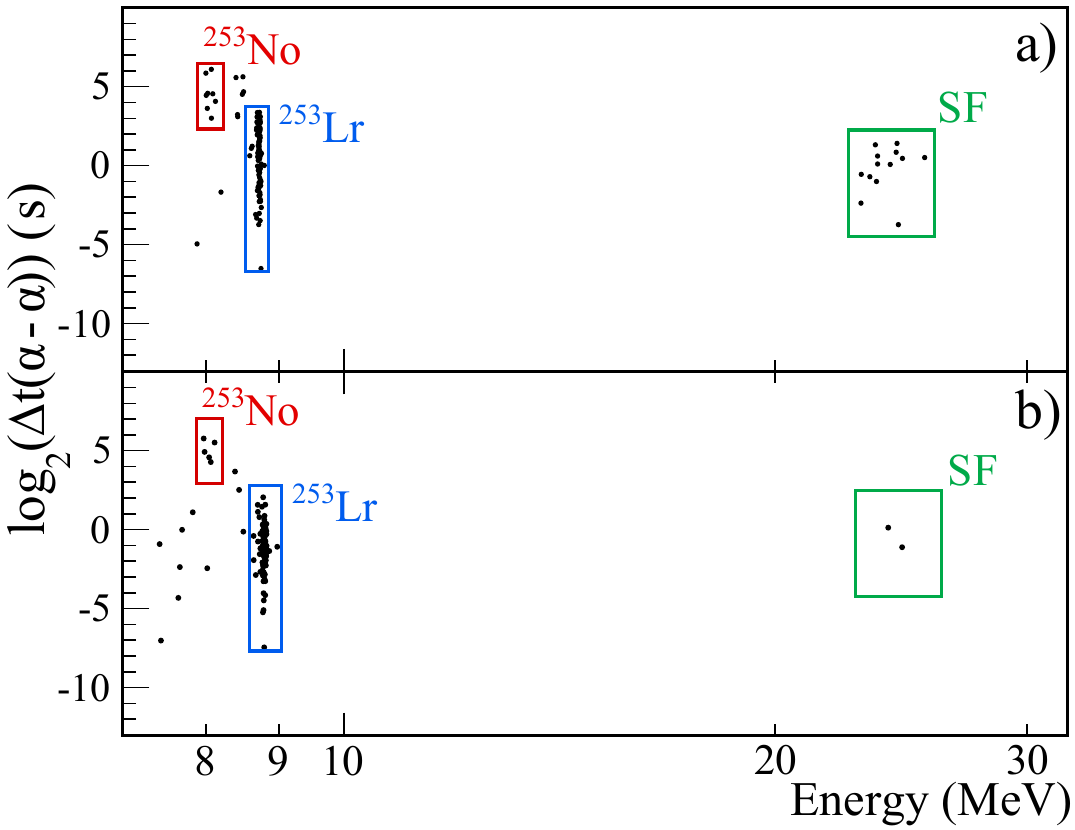}
      \caption{Decay time in log$_2$ of seconds as a function of the energy of the second-generation decays following $^{257}$Db: (a) low-spin state $\alpha$ decays and (b) high-spin state $\alpha$ decays, both with $\Delta t \leq 750$~ms. The red squares highlight the $\alpha$ decay of $^{253}$No after the EC of $^{253}$Lr, the green squares represent SF events of $^{253}$Lr, and the blue squares represent the $\alpha$ decays of $^{253}$Lr.}
      \label{BR_Lrge}
\end{figure}

The branching ratios for the two $\alpha$-decaying states in $^{253}$Lr can be estimated using the measured decays plotted in Fig.\ref{BR_Lrge}(a) for the high-spin state and Fig.\ref{BR_Lrge}(b) for the low-spin state, and are listed in Tab.~\ref{Recap_decay}. The blue squares represent the $\alpha$ decays of $^{253}$Lr, the red squares correspond to the $\alpha$ decays of $^{253}$No populated through Electron Capture (EC), and the green squares denote the spontaneous fission events (overflow of $\alpha$-decay range of the preamplifier~\cite{lopez2019}). The hindrance factor (HF) obtained for the $\alpha$ decays of $^{253}$Lr to $^{249}$Md indicate favoured decays in both cases.

\subsection{Decay properties of $^{249}$Md}

\begin{figure}[ht]
      \centering
     \includegraphics[width=1\linewidth]{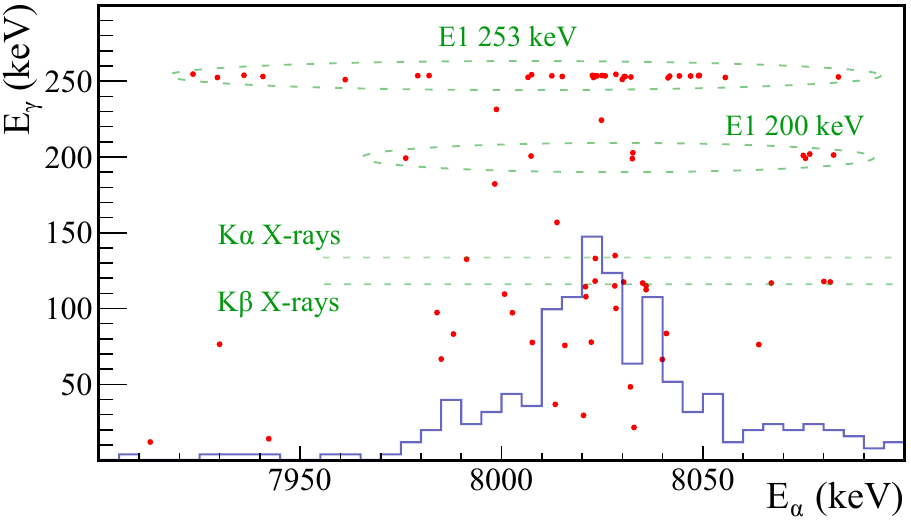}
      \caption{$\gamma$-ray energies as a function of coincident $^{249}$Md $\alpha$-decay energies. The green dashed circles highlight the known E1 transitions (253~keV and 200~keV) and the X-rays of $^{245}$Es. The blue spectrum is the spectrum of the $\alpha$-decay of $^{249}$Md feeding $^{245}$Es, extracted from Fig.~\ref{Eam_Ead}(b-c) (blue square).}
      \label{Trg_Es}
\end{figure}

The decay of $^{249}$Md into $^{245}$Es was studied in~\cite{249Md_gamma,hessberger2001decay,249Md_gamma,249Md_inbeam}. Two internal transitions following the decay of the 7/2$^{-}$[514] ground state of $^{249}$Md have been reported: an E1 transition of 253~keV connecting the 7/2$^{-}$[514] Nilsson state in $^{245}$Es to the 7/2$^{+}$[633] state, and an E1 transition of 200~keV connecting the 7/2$^{-}$[514] state to the 9/2$^{+}$ member of the 7/2$^{+}$[633] rotational band, as seen in Fig.\ref{Level}. In Fig.\ref{Trg_Es}, the 253~keV $\gamma$ ray is clearly observed in coincidence with $\alpha$-particle energies of 8022~keV, while the 200~keV transition is coincident with higher energy $\alpha$ particles due to summing with the coincident ICEs and accompanying atomic emissions.

\begin{figure}[ht]
      \centering
     \includegraphics[width=1\linewidth]{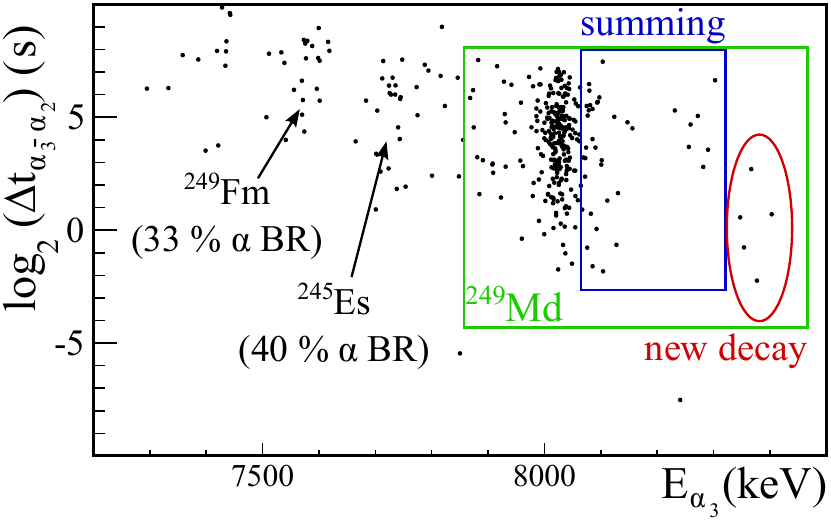}
      \caption{Decay time in log$_{2}$ of seconds for the third-degeneration decays in the $^{257}$Db chain. The green square corresponds to the $^{249}$Md decays. The blue square shows events for which there is a summation of $^{249}$Md $\alpha$ decays and internal transitions in $^{245}$Es. The red circle shows the newly reported $\alpha$ decay of $^{249}$Md.}
      \label{Tra_Ea3_Md}
\end{figure}

A previously unreported faster $\alpha$-decay of $^{249}$Md has also been observed. This new decay is highlighted in Fig.\ref{Tra_Ea3_Md} with the red circle, as well as in Fig.\ref{Eam_Ead}(b) within the purple dashed circle inside the $^{245}$Lr-$^{249}$Md/$^{249}$Fm red correlation box. In Fig.\ref{Eam_Ead}(b), the new decay is clearly only correlated to the low-energy $\alpha$ decay of $^{253}$Lr (E$_{\alpha} \leq$ 8748~keV), i.e., the decay of the 1/2$^{-}$ state. 

Figure~\ref{Tra_Ea3_Md} plots the logarithm of the decay time as a function of the decay energy for the third-generation decays. $^{245}$Es decays appear due to undetected escaping $\alpha$ particles, while the $\alpha$ decay of $^{249}$Fm originates from the EC of $^{249}$Md, $^{253}$No, or $^{257}$Db. The green square highlights all decays of $^{249}$Md, and the blue square corresponds to events where the $\alpha$ decay energy of $^{249}$Md is summed with the energies of internal transitions, Auger electrons, and L X-rays. The red circle highlights the newly observed $\sim$8350~keV $\alpha$ decay. Notably, this $\alpha$ decay branch is also observed in Fig.4 of~\cite{249Md_prod}, with the same energy and decay time, but was not interpreted.

The measured half-life of the 1/2$^-$ state in $^{249}$Md, extracted from Fig.~\ref{Tra_Ea3_Md}, is 1.1~$^{+0.6}_{-0.4}$~s, which is significantly faster than the 22.3(1.0)~s decay of the 7/2$^{-}$ ground state. By gating on the different $^{253}$Lr decays, one can extract $\alpha$-decay branching ratios for both states in $^{249}$Md, along with the corresponding hindrance factors (HF). There is no significant EC branch from the 1/2$^-$ state, as the ratio of the ground-state $\alpha$-decay branch to the EC branch is the same in both $^{253}$Lr $\alpha$-decay gates.

The HF obtained for the 1/2$^-$ decay of $^{249}$Md is $\sim$28, indicating a decay between two states of the same parity without a spin flip.


\section{Interpretation and Discussion}

\label{INTERPRETATION}

\subsection{Relative excitation energies of the high and low-spin states in the $^{257}$Db decay chains}

Based on the known level scheme of $^{245}$Es~\cite{249Md_gamma,hessberger2001decay} and the measured half-life of 1.1~s for the newly observed decay of the 1/2$^-$ state in $^{249}$Md, the detected 8350~keV $\alpha$ decay cannot feed the 7/2$^{-}$[514] state in $^{245}$Es, as this would place the 1/2$^-$ state at 331~keV above the ground state in $^{249}$Md. Its decay would then be dominated by an M3 transition to the 7/2$^-$ ground state, with a partial electromagnetic decay half-life of $\sim$8~ms.

The extracted HF for the 8350~keV decay is also too small to allow for a direct decay to the 7/2$^{+}$[633] state of $^{245}$Es, as such a decay would involve a parity change. It is however compatible with a L~=~2 decay to the 3/2$~{-}$[521] ground state, the L~=~0 component being much more hindered due to spin flip. Population to the 5/2$^{-}$ member of the band at an excitation energy of $\sim$30 keV, deduced from systematics, may also contribute to the transition.

Geant4 simulations of the decay (as described in~\cite{chakma2020gamma}) were used to check the decay scheme of Fig.\ref{249Md_level}. The simulations include all the radiative and non-radiative processes that accompany the internal conversion process. The implantation depth of $^{257}$Db ions, which is a crucial factor for the proper modeling of summing events, was determined according to the procedure described in~\cite{chakma2020gamma} by adjusting the average and standard deviation of a Gaussian implantation profile to reproduce the energy depositions of the $^{257}$Db $\alpha$ particles that escape from the DSSD. 

\begin{figure}[ht]
      \centering
     \includegraphics[width=1\linewidth]{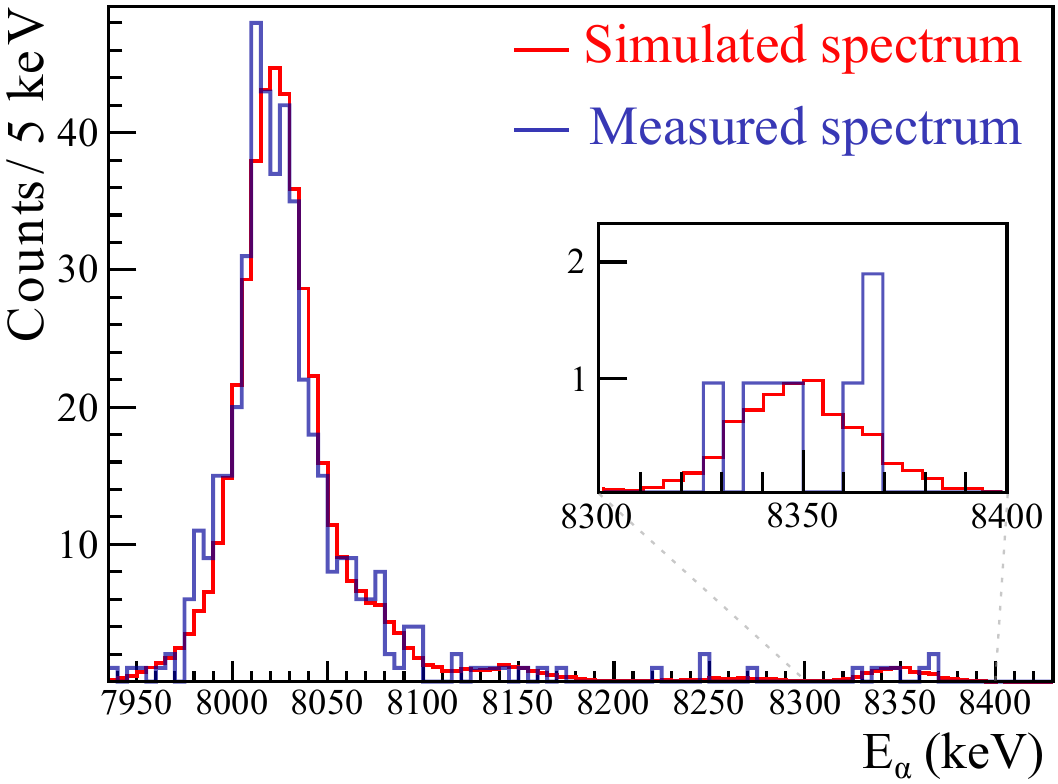}
      \caption{Experimental $\alpha$-decay spectrum of $^{249}$Md (blue) extracted from Fig.\ref{Tra_Ea3_Md}. The simulated spectrum (red) was obtained using an $\alpha$-decay energy of 8350~keV for the decay of the 1/2$^-$ state. For the decay of the 7/2$^{-}$ ground state of $^{249}$Md, the simulations include a 20~\% branch to the 9/2$^{-}$ member of the 7/2$^{-}$ rotational band in $^{245}$Es. The inset illustrates the sensitivity of the simulation when reproducing the newly detected decay from the 1/2$^-$ state in $^{249}$Md.}
      \label{sim249}
\end{figure}

\begin{figure}[ht]
      \centering
     \includegraphics[width=1\linewidth]{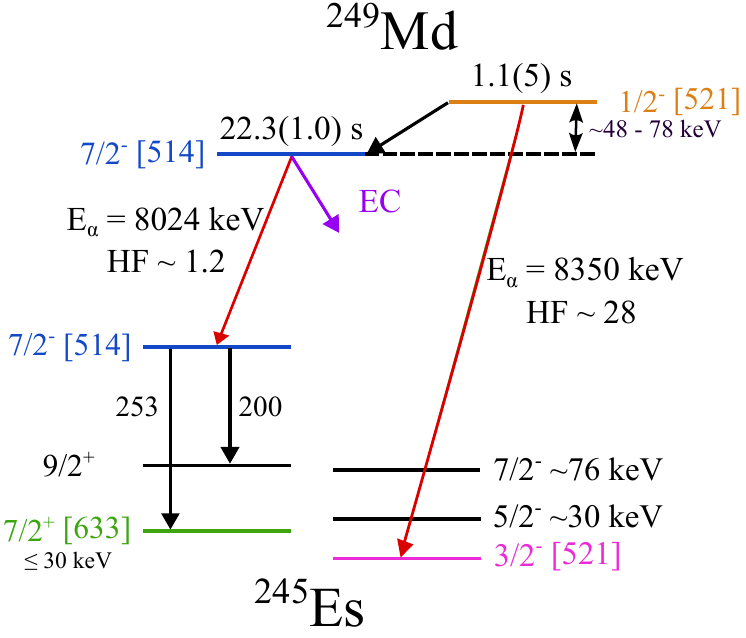}
      \caption{Decay scheme of $^{249}$Md obtained from this work.}
      \label{249Md_level}
\end{figure}

The results of the simulation of the newly observed decay of the 1/2$^-$ state in $^{249}$Md can be seen in Fig.\ref{sim249}. The potential feeding of the 5/2$^{-}$ member of the ground state band with a decay energy of 8320~keV was not included as the measured statistics was not enough to evaluate its contribution.

This interpretation of the decay data of $^{249}$Md differs from that of $^{247}$Md~\cite{247Md}, where the hindrance factor for the $\alpha$ decay of the 1/2$^-$ isomeric state was found to be 3.6, indicating a favoured transition between states of the same configuration. However, this hindrance was obtained under the assumption of no internal decay branch, even though such a branch is expected, as an M3 transition of 153~keV to the ground state should be able to compete with a 0.23~s $\alpha$ decay. The presence of this internal branch appears to be confirmed by the genetic correlations observed in the discovery paper of $^{251}$Lr~\cite{251Lr}, which shows that all the $\alpha$ particles emitted by $^{251}$Lr are correlated with the ground-state decay of $^{247}$Md.

The excitation energy of the 7/2$^{+}$[633] state in $^{245}$Es is not known, but it must be below the 5/2$^{-}$ member of the ground-state band, making it isomeric, since no L X-rays have been detected. In addition, the width and shape of the main $\alpha$ peak are well reproduced by the simulations, which do not include a subsequent 30~keV M1 decay. The obtained decay scheme of $^{249}$Md is shown in Fig.\ref{249Md_level}.

\begin{figure}[ht]
      \centering
     \includegraphics[width=1\linewidth]{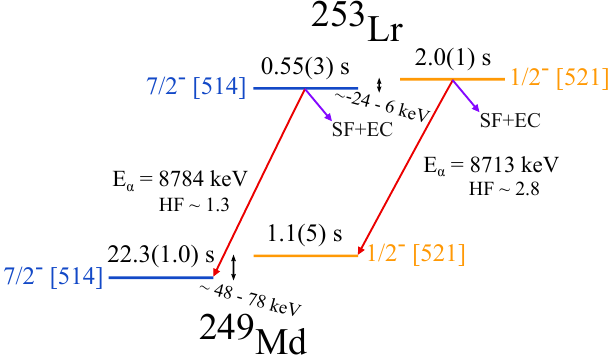}
      \caption{Decay scheme of $^{253}$Lr obtained from this work.}
      \label{253Lr_level}
\end{figure}

The excitation energy of the 1/2$^-$ state in $^{249}$Md lies between 48 and 78~keV. An M3 internal decay to the ground state would have a partial half-life of $\sim$0.3~s, which is consistent with the observed half-life and the deduced internal decay branching ratio. With the excitation energy of the 1/2$^-$ isomer in $^{249}$Md established, the measured decay properties of $^{253}$Lr allow for determining its relative position with respect to the 7/2$^-$ state in $^{253}$Lr, as shown in Fig.~\ref{253Lr_level}.

The 7/2$^-$ and 1/2$^-$ states are found to be nearly degenerate, which explains why no internal decay was observed, unlike in the case of $^{255}$Lr, where the 1/2$^-$ state at $\sim$30~keV above the ground state is known to have a large internal decay branch. A lowering of the 1/2$^-$ state relative to the 7/2$^-$ state as neutrons are removed from the system is consistent with the inversion of the ordering of the 7/2$^{-}$[514] and 1/2$^{-}$[521] states in $^{251}$Lr compared to $^{255}$Lr~\cite{251Lr}.

Finally, in the case of $^{257}$Db, the energy positions of the two $\alpha$-decaying states can be estimated by taking the endpoint of the low-energy group of $\alpha$-particle energies (minus the experimental FWHM of $\sim$31~keV) as the maximum energy released in the decay of the high-spin state. This leads to an excitation energy for the 1/2$^-$ state between 59 and 89~keV above the high-spin ground state.

\subsection{Nature of the high-spin ground state in $^{257}$Db} 

In F.P. He{\ss}berger \textit{et al.}~\cite{hessberger2001decay}, the low-energy double-humped alpha-particle energy spectrum is interpreted as arising from a 8964~keV decay branch into the 9/2$^{+}$[624] level of $^{253}$Lr and from a 9074~keV decay into the 7/2$^{-}$[514] ground state $^{253}$Lr (both branches being of equal relative intensity)

In addition, as noted by F.P. He{\ss}berger \textit{et al.}~\cite{hessberger2001decay}, no E1 transition is observed between the 9/2$^{+}$ state and the 7/2$^-$ ground state or other members of the ground-state band. Furthermore, they also noted that the two peaks of the low-energy double-humped $\alpha$-particle structure of $^{257}$Db had larger widths than expected. They concluded that this was due to summing between $\alpha$-particle and internal conversion electron (+Auger) energies arising from M1 decays within the ground state band.

However, the decay from the 9/2$^{+}$[624] state of $^{257}$Db to the negative-parity 7/2$^-$[514] ground state of $^{253}$Lr would have a hindrance factor greater than 1000, due to the required parity and spin flip. This would make such a decay highly unlikely to be observed with the intensity suggested by the earlier interpretation. 
Therefore, we reinterpret the decay scheme from F.P. He{\ss}berger \textit{et al.}~\cite{hessberger2001decay}, proposing that only one decay occurs from the 9/2$^{+}$[624] state in $^{257}$Db to its equivalent in $^{253}$Lr.

\begin{figure}[ht]
      \centering
     \includegraphics[width=1\linewidth]{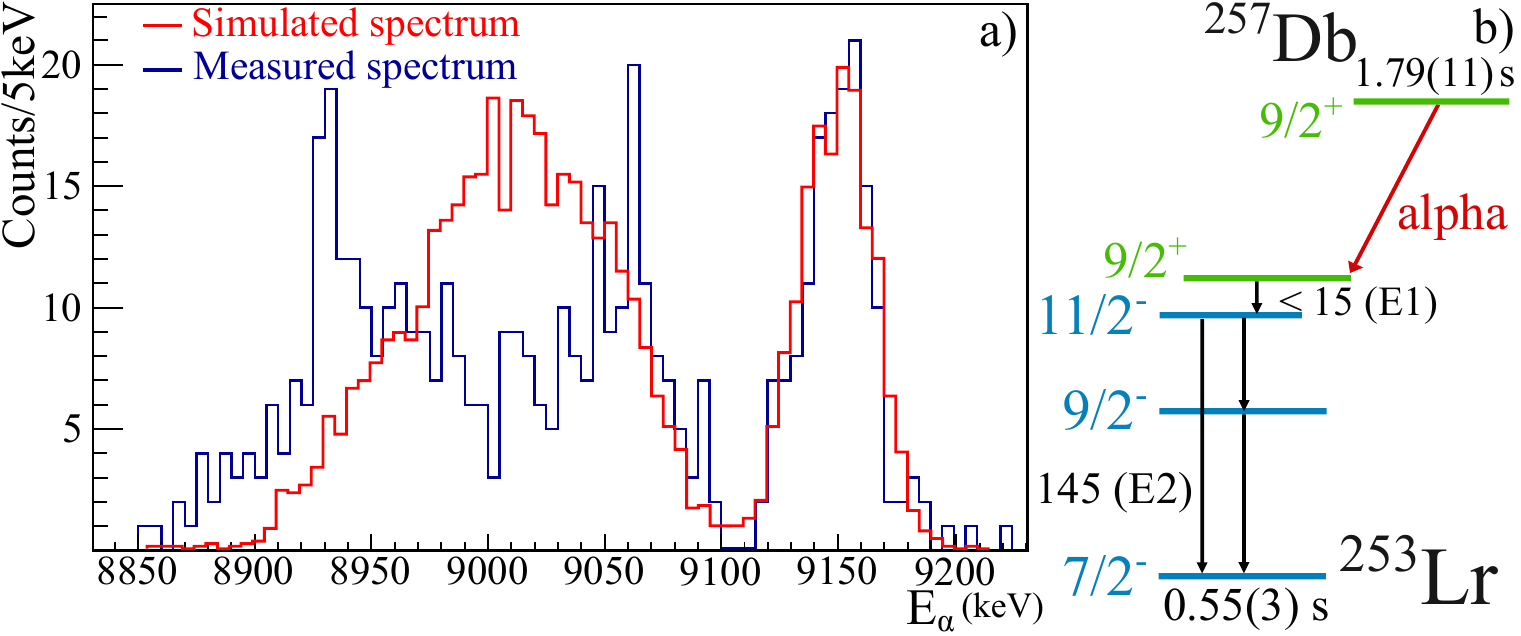}
      \caption{(a) Comparison between the measured $\alpha$-decay energy spectrum of $^{257}$Db (blue) and the simulated one (red) based on the previously proposed level scheme~\cite{hessberger2001decay,streicher2006synthesis}. (b) Modified level scheme based on the proposed one from~\cite{hessberger2001decay,streicher2006synthesis} used for the simulation.}
      \label{253Lr_E1}
\end{figure}

Furthermore, given the observation of the highly converted 145~keV transition, we propose that the 9/2$^{+}$ state preferentially decays to the 11/2$^{-}$ member of the 7/2$^-$[514] ground state. The E2 decay from this 11/2$^{-}$ state to the ground state could, in principle, account for the observed counts at 145~keV. Moreover, the E1 decay from the 9/2$^{+}$ to the 11/2$^{-}$ member of the ground-state band must be $<$15~keV given our detection efficiency, which is $>$10--15~\% above 15--20~keV~\cite{chakma2020gamma}. This scenario is illustrated in Fig.\ref{253Lr_E1}(b), with the corresponding simulated spectrum displayed in Fig.\ref{253Lr_E1}(a).

As seen from Fig.\ref{253Lr_E1}(a), the decay scenario depcited in Fig.\ref{253Lr_E1}(b) cannot account for the double-humped structure observed in the high-spin state decay. The Geant4 simulations shown in Fig.\ref{253Lr_E1}(a) were conducted under the assumption that the 145~keV transition corresponds to the 11/2$^-\rightarrow$7/2$^-$ intraband E2 transition of the 7/2$^-$ ground-state band, using the same gyromagnetic factor for the 7/2$^-$[514] configuration as determined experimentally in $^{251}$Md~\cite{249Md_inbeam}. Under this scenario, the resulting $\alpha$ spectrum exhibits a single hump. 
This configuration arises due to the gyromagnetic factor of the ground-state rotational band (g$_K$ = 0.69), which favours small M1 decays over the crossover E2 transition.
If emitted in the forward direction, the M1 transitions would be absorbed by the DSSD, just like the E2 transitions would. However, if emitted backward, the E2 transitions have a much greater probability of escaping the DSSD undetected, which results in them not contributing to the detected $\alpha$-particle energy.
To replicate the observed “double-hump” $\alpha$ spectrum in Fig.\ref{Ea_fit}, a $\sim$145~keV M1 or E2 transition connecting states from different rotational bands is necessary.

A possibility could involve the 7/2$^{+}$[633] state in $^{253}$Lr. If this state were located above the 9/2$^{+}$[624] state in $^{253}$Lr, the $\alpha$-decay branch feeding the 7/2$^{+}$[633] state would be too small to produce the double-hump structure. Moreover, the E1 decay from the 7/2$^{+}$ state to the ground-state band would have been visible. On the other hand, if the 7/2$^{+}$[633] state were located below the 9/2$^{+}$[624] state, the double-hump structure could in principle be reproduced with a decay to the 7/2$^{+}$ band. However, the E1 decay from the 9/2$^{+}$ state to the 7/2$^-$ and 9/2$^-$ members of the ground-state band would have beeen observed. Finally, if the 9/2$^{+}$ state only decayed via a small unobserved E1 decay to the 11/2$^{-}$ member of the ground-state band, this would be equivalent to the scenario of Fig.~\ref{253Lr_E1}.

As a side note, the 7/2$^{+}$[633] state lies below $Z$~=~100, as shown in Fig.~\ref{249Md_level}, and is close to the ground state of $^{245}$Es. It is unlikely to be near the Fermi surface in $^{253}$Lr.

Thus, whichever way one tries to accomodate the 9/2$^{+}$[624] state in the decay of the high spin state of the $^{257}$Db, the fundamental issue remains: the 9/2$^{+}$[624] and 7/2$^{-}$[514] orbitals have opposite parities and spin projections. This mismatch explains why no scenario involving a 9/2$^{+}$[624] ground state in $^{257}$Db and a 7/2$^{-}$[514] ground state in $^{253}$Lr can satisfactorily account for both the observed $\alpha$-particle energy spectrum and internal transitions. 

\begin{figure}[ht]
      \centering
     \includegraphics[width=0.95\linewidth]{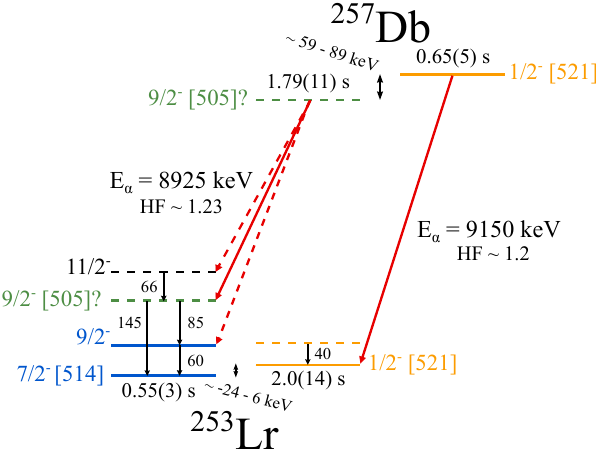}
      \caption{Proposed decay scheme of $^{257}$Db extracted from this work and analysis/discusssion.}
      \label{257Db_level}
\end{figure}

The scenario proposed here, depicted in Fig.~\ref{257Db_level}, assumes that the previous 7/2$^-$ assignments for the ground states of $^{253}$Lr and $^{249}$Md is correct. Additionally, the newly observed high-K isomer decays confirm that the ground state of $^{257}$Db must be a high-spin state. It must also have a high enough spin to explain the half-life of the 1/2$^{-}$[521] state, which rules out the 5/2$^{-}$[512] state. Based on these required characteristics, we propose that the ground state of $^{257}$Db corresponds to the 9/2$^{-}$[505] Nilsson state, which originates from the same $h_{9/2}$ shell as the 7/2$^{-}$[514] state.

Figure~\ref{257Db_level} presents the proposed decay scheme of $^{257}$Db. The primary $\alpha$-decay branch from the 9/2$^{-}$[505] state in $^{257}$Db populates the corresponding 9/2$^{-}$[505] state in $^{253}$Lr, located 145~keV above the 7/2$^{-}$[514] ground state. To reproduce the observed two-humped decay structure, additional lower-intensity branches were considered: an 18.8~\% branch feeding the 11/2$^-$ members of rotational band built on top the the 9/2$^-$[505] state and a 5~\% branch populating the 9/2$^-$ member of the ground-state band (see Table~\ref{Recap_decay}).

\begin{figure}[ht!]
      \centering
     \includegraphics[width=1\linewidth]{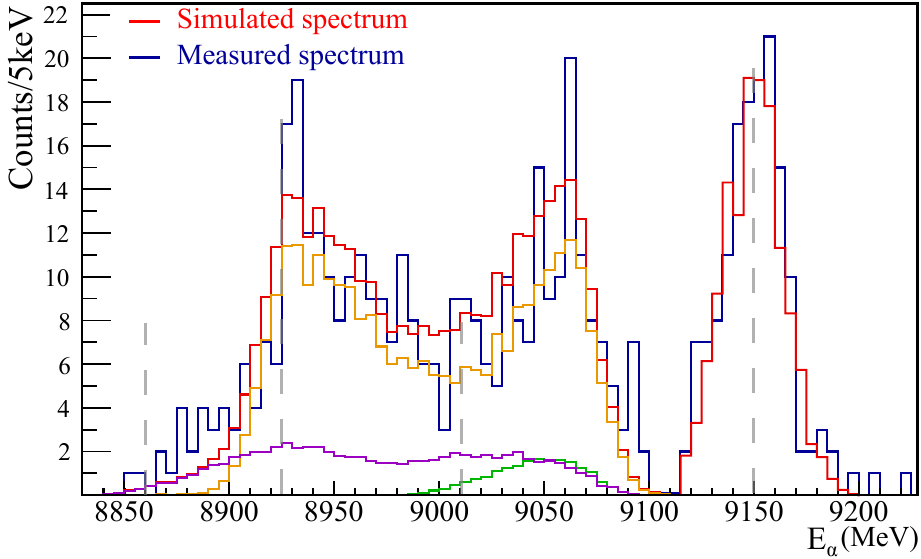}
      \caption{Comparison between the measured $\alpha$-decay energy spectrum of $^{257}$Db (red) and the simulated one (blue) based on the decay scheme shown in Fig.~\ref{257Db_level}. The other colored spectra represent the different contributions to the high-spin alpha decay shown in Fig.~\ref{257Db_level}: In orange the alpha decay to the 9/2$^-$[505] band head, in purple to the 11/2$^-$ member of the 9/2$^-$[505] band, and in green the contribution of the 9/2$^-$ member of the g.s band.}
      \label{Compa_simu}
\end{figure}

Figure~\ref{Compa_simu} highlights the reproduction of the data through Geant4 simulation with this scenario; orange is the contribution of the decay to the 9/2$^-$[505] band head, green is to the 9/2$^-$ member of the ground-state band, and purple is to the 11/2$^-$ member of the 9/2$^-$[505] band. The extracted intensities yield a HF of the order of 1.2 and 34 for the decays to the 9/2$^-$[505] band head and to the 9/2$^-$ member of the ground-state band. For the decay to the 11/2$^-$ member of the 9/2$^-$[505] head band, the intensities yield a HF of the order of 3.1.

This revised decay scenario also explains the observed coincidence between $\alpha$ decays and the $\gamma$ rays or internal conversion electrons (ICEs) emitted by $^{253}$Lr. As discussed in Section~\ref{RESULTS}, the measured conversion coefficient for the 145~keV transition is 21$^{+28}_{-11}$. Although this value has a large uncertainty, it remains consistent within less than 2$\sigma$ with the expected M1 or E2 LMN+ conversion coefficients of 4.95 and 6.4, respectively.

\subsection{Discussion}

This interpretation of the data raises two issues. The first one is the position of the previously suggested 9/2+[624] state in $^{257}$Db (see Fig.~\ref{Level}), which has been only very tentatively assigned to levels in $^{251}$Es and $^{249}$Bk~\cite{ahmad1978,ahmad2005} as well as $^{255}$Lr~\cite{jeppesen2009}. The state is also involved in the $\pi^2(1/2^-[521]\otimes 9/2^+[624])_{5^-} \otimes \nu 11/2^-[725]$ assignment of the 21/2$^+$ high-K isomer observed in $^{257}$Rf~\cite{257Rf}, which was preferred over the $\nu^2(9/2^-[734]\otimes 1/2^+[620])_{5^-}\otimes \nu 11/2^-[725]$ assignment on the basis of theoretical calculations. The 9/2$^+$[624] orbital is also thought to be active in the 8$^-$ high-K state in $^{254}$No~\cite{tandel2006k, herzberg2001spectroscopy,clark2010high}, though a two-quasi-neutron configuration cannot be ruled out.

The UNEDF1 functional~\cite{shi2014rotational}, predicts that the 9/2$^{+}$[624] Nilsson state, stemming from the $i_{13/2}$ intruder shell, should lie further away from the Fermi surface than in the Woods Saxon picture (the difference is illustrated in Figs. 7 a) and b) of Y. Shi $et\ al.$~\cite{shi2014rotational}).
One possible explanation for the the closeness in energy of the 7/2$^{-}$ and 9/2$^-$ states may be the onset of higher-multipolarity deformation changes in this region. Indeed both macroscopic-microscopic, and purely microscopic, self-consistent approaches~\cite{sobiczewski2007description} predict a local minimum in $\beta_{6}$ centered around $Z$~=~104 and $N$~=~152 and a steady decrease of $\beta_{4}$ towards negative values as $Z$ and $N$ increase respectively from $Z$~=~100 to 114 and $N$~=~150 to 164. In Parkhomenko and Sobiczewski~\cite{parkhomenko2006single}, but also Cwiok~\cite{CWIOK1994356}, $\beta_{4}$ is predicted to vary from 0.04 in $^{249}$Md to 0.016--0.017 in $^{253}$Lr to -0.006 in $^{257}$Db. As illustrated in Fig. 9 of Chasman $et\ al.$~\cite{RevModPhys.49.833}, the 9/2$^-$[505] single-proton state becomes more bound as $\beta_{4}$ decreases, gaining ~150 keV of binding energy when $\beta_{4}$ changes from $\beta_{4}$=0.04 to $\beta_{4}$=0, while the opposite behaviour is seen for the 7/2$^{-}$[514] and 1/2$^-$[521] proton states, the 7/2$^{-}$[514] losing $\sim$400 keV of binding energy. These deformation changes are also thought to be responsible for the trend of the neutron $j_{15/2}$ 11/2$^{-}$[725] state in $N$~=~153 isotones~\cite{PhysRevC.110.054310} as predicted by~\cite{CWIOK1994356,N153} as well as the level inversion in Lr isotopes~\cite{251Lr}. It is to be noted, however, that the behaviours as a function of $\beta_{4}$ of the proton states of interest calculated by Cwiok~\cite{CWIOK1994356} are opposite to the ones of Chasman~\cite{RevModPhys.49.833} and therefore do not support the argument of a lowering of the 9/2$^-$[505] state with respect to the 7/2$^{-}$[514] state due to higher-multipolarity deformation changes.

 
\section{CONCLUSION} 

In conclusion, a new study of the decay properties of $^{257}$Db has revealed the presence of a second isomeric state with an half-life of 0.71~$^{+0.29}_{-0.18}$~ms, most likely a high-K isomer since it lies at high excitation energy and solely decays to the high-spin ground state of $^{257}$Db. By following the $\alpha$ decay chains of the high-spin and low-spin states in $^{257}$Db all the way down to $^{245}$Es, their relative excitation energies could be narrowed down to within 30~keV in $^{249}$Md, $^{253}$Lr and $^{257}$Db. 

In order to explain the observed double-humped structure of the $\alpha$-decay spectrum of the high-spin ground state in $^{257}$Db as well as the presence of a low-lying low-spin $\alpha$-decaying state in $^{257}$Db, a radical change in the structure of the ground state of $^{257}$Db has been proposed, which involves the 9/2$^{-}$[505] state, originating from the same $h_{9/2}$ shell as the 7/2$^-$[514] ground state of $^{253}$Lr. The 9/2$^-$[505] state is the only one found to meet the parity and spin criteria to account for all the experimental observations. The presence of this state at the Fermi surface as well as the closeness to the 7/2$^-$[514] state are difficult to explain, though some arguments have been put forward, in particular a relative positioning of spherical high-$j$ proton shells more in line with density functional theory predictions and high-mulipolarity deformation effects. If confirmed, this would indicate that the experimental spectrum of proton states in $^{257}$Db and $^{253}$Lr is not properly described by non-self consistent calculations, which could in turn signify the importance of charge density effects. More data in odd $Z$ nuclei are clearly needed to understand the sequence and nature of proton states above $Z$~=~100.



\section*{Acknowledgements}

The authors address a special thanks to the U400 cyclotron crew, to the ion-source team and to the Dubna Chemists who prepared the $^{209}$Bi target. 
Recent upgrade of the separator SHELS (modernized VASSILISSA) and detection system GABRIELA the CLODETTE germanium clover detector and fore coming digital electronics was partially financed by the French ``Agence Nationale de la Recherche” (ANR) under the Contracts No. ANR-06-BLAN-0034-01 and ANR-12-BS05-0013. The GABRIELA project is jointly funded by JINR (Russia) and IN2P3/CNRS (France). Work at FLNR was performed partially under the financial support of the Russian Foundation on Basic Research (RFBR), Contracts No. 08-02-00116, 17-02-00867 and 18-52-15004.

%
%
%






\bibliographystyle{elsarticle-num}
\bibliography{Biblio}

\begin{thebibliography}{10}
\expandafter\ifx\csname url\endcsname\relax
  \def\url#1{\texttt{#1}}\fi
\expandafter\ifx\csname urlprefix\endcsname\relax\def\urlprefix{URL }\fi
\expandafter\ifx\csname href\endcsname\relax
  \def\href#1#2{#2} \def\path#1{#1}\fi

\bibitem{SRILAC}
H.~Sakai, H.~Haba, K.~Morimoto, N.~Sakamoto, Facility upgrade for
  superheavy-element research at $\mathrm{RIKEN}$, The European Physical
  Journal A-Hadrons and Nuclei 58~(12) (2022) 238.

\bibitem{SHE_factory}
S.~Dmitriev, M.~Itkis, Y.~Oganessian, Status and perspectives of the {D}ubna
  superheavy element factory 131 (2016) 08001.

\bibitem{SHANS2}
S.~Xu, Z.~Zhang, Z.~Gan, M.~Huang, L.~Ma, J.~Wang, M.~Zhang, H.~Yang, C.~Yang,
  Z.~Zhao, et~al., A gas-filled recoil separator, {SHANS2}, at the {C}hina
  accelerator facility for superheavy elements, Nuclear Instruments and Methods
  in Physics Research Section A: Accelerators, Spectrometers, Detectors and
  Associated Equipment 1050 (2023) 168113.

\bibitem{gates2024towards}
J.~M. Gates, R.~Orford, D.~Rudolph, C.~Appleton, B.~M. Barrios, J.~Y. Benitez,
  M.~Bordeau, W.~Botha, C.~M. Campbell, J.~Chadderton, et~al., Toward the
  discovery of new elements: Production of {L}ivermorium ($\mathrm{Z}=116$)
  with $^{50}\mathrm{Ti}$, Physical Review Letter 133 (2024) 172502.

\bibitem{GABRIELA}
K.~Hauschild, A.~Yeremin, O.~Dorvaux, A.~Lopez-Martens, A.~Belozerov,
  C.~Briançon, M.~Chelnokov, V.~Chepigin, S.~Garcia-Santamaria, V.~Gorshkov,
  et~al., {GABRIELA}: A new detector array for $\gamma$-ray and conversion
  electron spectroscopy of transfermium elements, Nuclear Instruments and
  Methods in Physics Research Section A: Accelerators, Spectrometers, Detectors
  and Associated Equipment 560~(2) (2006) 388--394.

\bibitem{SHELS}
A.~Popeko, A.~Yeremin, O.~Malyshev, V.~Chepigin, A.~Isaev, Y.~Popov,
  A.~Svirikhin, K.~Haushild, A.~Lopez-Martens, K.~Rezynkina, O.~Dorvaux,
  Separator for {H}eavy {EL}ement {S}pectroscopy - velocity filter {SHELS},
  Nuclear Instruments and Methods in Physics Research Section B: Beam
  Interactions with Materials and Atoms 376 (2016) 140--143, proceedings of the
  XVIIth International Conference on Electromagnetic Isotope Separators and
  Related Topics (EMIS2015), Grand Rapids, MI, U.S.A., 11-15 May 2015.

\bibitem{hessberger2001decay}
F.~He{\ss}berger, S.~Hofmann, D.~Ackermann, V.~Ninov, M.~Leino,
  G.~M{\"u}nzenberg, S.~Saro, A.~Lavrentev, A.~Popeko, A.~Yeremin, et~al.,
  Decay properties of neutron-deficient isotopes $^{256,257}${D}b,
  $^{255}\mathrm{Rf}$, $^{252,253}\mathrm{Lr}$, The European Physical Journal
  A-Hadrons and Nuclei 12 (2001) 57--67.

\bibitem{gates2008comparison}
J.~M. Gates, S.~L. Nelson, K.~E. Gregorich, I.~Dragojevi{\'c}, C.~E.
  D{\"u}llmann, P.~A. Ellison, C.~M. Folden~III, M.~A. Garcia, L.~Stavsetra,
  R.~Sudowe, et~al., Comparison of reactions for the production of
  $\mathrm{Db}^{258,257}$: $^{208}\mathrm{Pb}$ ($^{51}\mathrm{V}$, xn) and
  $^{209}\mathrm{Bi}$ ($^{50}\mathrm{Ti}$, xn), Physical Review C 78~(3) (2008)
  034604.

\bibitem{streicher2006synthesis}
P.~Brionnet, Etude des états isomères des noyaux superlourds : cas des noyaux
  $^{257}\mathrm{Db}$ et $^{243}\mathrm{Lr}$, Ph.D. thesis (2017).

\bibitem{pbrionnet}
B.~Streicher, Synthesis and spectroscopic properties of transfermium isotopes
  with $\mathrm{Z}$= 105, 106 and 107, Ph.D. thesis (2006).

\bibitem{ECR}
S.~Bogomolov, V.~Bekhterev, A.~Efremov, B.~Gikal, G.~Gulbekian, Y.~Kostukhov,
  A.~Lebedev, V.~Loginov, N.~Yazvitsky, Recent development in {ECR} ion sources
  at {FLNR} {JINR}, in: Proc. Russian Particle Accelerator Conf. RUPAC, 2012,
  pp. 203--207.

\bibitem{MIVOC}
J.~Rubert, J.~Piot, Z.~Asfari, B.~Gall, J.~Ärje, O.~Dorvaux, P.~Greenlees,
  H.~Koivisto, A.~Ouadi, R.~Seppälä, First intense isotopic titanium-50 beam
  using {MIVOC} method, Nuclear Instruments and Methods in Physics Research
  Section B: Beam Interactions with Materials and Atoms 276 (2012) 33--37.

\bibitem{U-400}
V.~B. Kutner, S.~L. Bogomolov, A.~A. Efremov, A.~N. Lebedev, V.~Y. Lebedev,
  V.~N. Loginov, A.~B. Yakushev, N.~Y. Yazvitsky, {Production of intense
  $^{48}${C}a ion beam at the {U-400} cyclotron}, Review of Scientific
  Instruments 71~(2) (2000) 860--862.

\bibitem{chakma2020gamma}
R.~Chakma, K.~Hauschild, A.~Lopez-Martens, A.~Yeremin, O.~Malyshev, A.~Popeko,
  Y.~A. Popov, A.~Svirikhin, V.~Chepigin, O.~Dorvaux, et~al., Gamma and
  conversion electron spectroscopy using {GABRIELA}, The European Physical
  Journal A-Hadrons and Nuclei 56 (2020) 1--10.

\bibitem{lopez2019}
A.~Lopez-Martens, A.~Yeremin, M.~Tezekbayeva, Z.~Asfari, P.~Brionnet,
  O.~Dorvaux, B.~Gall, K.~Hauschild, D.~Ackermann, L.~Caceres, et~al.,
  Measurement of proton-evaporation rates in fusion reactions leading to
  transfermium nuclei, Physics Letters B 795 (2019) 271--276.

\bibitem{249Md_gamma}
F.~He{\ss}berger, S.~Antalic, B.~Streicher, S.~Hofmann, D.~Ackermann,
  B.~Kindler, I.~Kojouharov, P.~Kuusiniemi, M.~Leino, B.~Lommel, et~al., Energy
  systematics of low-lying {N}ilsson levels in odd-mass einsteinium isotopes,
  The European Physical Journal A-Hadrons and Nuclei 26 (2005) 233--239.

\bibitem{249Md_inbeam}
R.~Briselet, C.~Theisen, B.~Sulignano, M.~Airiau, K.~Auranen, D.~M. Cox,
  F.~D\'echery, A.~Drouart, Z.~Favier, B.~Gall, et~al., In-beam
  $\ensuremath{\gamma}$-ray and electron spectroscopy of
  $^{249,251}\mathrm{Md}$, Physical Review C 102 (2020) 014307.

\bibitem{249Md_prod}
R.~Briselet, C.~Theisen, M.~Vandebrouck, A.~Marchix, M.~Airiau, K.~Auranen,
  H.~Badran, D.~Boilley, T.~Calverley, D.~Cox, et~al., Production cross section
  and decay study of $^{243}\mathrm{Es}$ and $^{249}\mathrm{Md}$, Physical
  Review C 99 (2019) 024614.

\bibitem{247Md}
F.~He{\ss}berger, S.~Antalic, F.~Giacoppo, B.~Andel, D.~Ackermann, M.~Block,
  S.~Heinz, J.~Khuyagbaatar, I.~Kojouharov, M.~Venhart, Alpha-gamma decay
  studies of $^{247}\mathrm{Md}$, The European Physical Journal A-Hadrons and
  Nuclei 58~(1) (2022) 11.

\bibitem{251Lr}
T.~Huang, D.~Seweryniak, B.~B. Back, P.~C. Bender, M.~P. Carpenter,
  P.~Chowdhury, R.~M. Clark, P.~A. Copp, X.-T. He, R.~D. Herzberg, D.~E.~M.
  Hoff, H.~Jayatissa, T.~L. Khoo, F.~G. Kondev, G.~Morgan, C.~Morse,
  A.~Korichi, T.~Lauritsen, C.~M\"uller-Gatermann, et~al., Discovery of the new
  isotope $^{251}\mathrm{Lr}$: {I}mpact of the hexacontetrapole deformation on
  single-proton orbital energies near the $\mathrm{Z}=100$ deformed shell gap,
  Physical Review C 106 (2022) L061301.

\bibitem{ahmad1978}
I.~Ahmad, R.~Sjoblom, A.~Friedman, S.~Yates, Proton states in the $\mathrm{Z}$=
  99 nucleus $^{251}\mathrm{Es}$ excited by $^{251}\mathrm{Fm}$ electron
  capture decay and $^{250}\mathrm{Cf}$ ($\alpha$, t) reaction, Physical Review
  C 17~(6) (1978) 2163.

\bibitem{ahmad2005}
I.~Ahmad, F.~G. Kondev, E.~F. Moore, M.~P. Carpenter, R.~R. Chasman, J.~P.
  Greene, R.~V.~F. Janssens, T.~Lauritsen, C.~J. Lister, D.~Seweryniak, R.~W.
  Hoff, J.~E. Evans, R.~W. Lougheed, C.~E. Porter, L.~K. Felker, Energy levels
  of $^{249}\mathrm{Bk}$ populated in the \ensuremath{\alpha} decay of
  ${}_{99}^{253}\mathrm{Es}$ and ${\ensuremath{\beta}}^{\ensuremath{-}}$ decay
  of ${}_{96}^{249}\mathrm{Cm}$, Phys. Rev. C 71 (2005) 054305.

\bibitem{jeppesen2009}
H.~Jeppesen, R.~Clark, K.~Gregorich, A.~Afanasjev, M.~Ali, J.~Allmond,
  C.~Beausang, M.~Cromaz, M.~Deleplanque, I.~Dragojevi{\'c}, et~al.,
  High-$\mathrm{K}$ multi-quasiparticle states and rotational bands in
  $^{253}_{103}\mathrm{Lr}$, Physical Review C 80~(3) (2009) 034324.

\bibitem{257Rf}
J.~Rissanen, R.~Clark, K.~Gregorich, J.~Gates, C.~Campbell, H.~Crawford,
  M.~Cromaz, N.~Esker, P.~Fallon, U.~Forsberg, et~al., Decay of the
  high-$\mathrm{K}$ isomeric state to a rotational band in $^{257}\mathrm{Rf}$,
  Physical Review C 88~(4) (2013) 044313.

\bibitem{tandel2006k}
S.~Tandel, T.~Khoo, D.~Seweryniak, G.~Mukherjee, I.~Ahmad, B.~Back,
  R.~Blinstrup, M.~Carpenter, J.~Chapman, P.~Chowdhury, et~al., K isomers in
  $\mathrm{No}^{254}$: Probing single-particle energies and pairing strengths
  in the heaviest nuclei, Physical Review Letters 97~(8) (2006) 082502.

\bibitem{herzberg2001spectroscopy}
R.-D. Herzberg, N.~Amzal, F.~Becker, P.~Butler, A.~Chewter, J.~Cocks,
  O.~Dorvaux, K.~Eskola, J.~Gerl, P.~Greenlees, et~al., Spectroscopy of
  transfermium nuclei: $^{252}_{102}\mathrm{No}$, Physical Review C 65~(1)
  (2001) 014303.

\bibitem{clark2010high}
R.~Clark, K.~Gregorich, J.~Berryman, M.~Ali, J.~Allmond, C.~Beausang,
  M.~Cromaz, M.~Deleplanque, I.~Dragojevi{\'c}, J.~Dvorak, et~al., High-{K}
  multi-quasiparticle states in $^{254}\mathrm{No}$, Physics Letters B 690~(1)
  (2010) 19--24.

\bibitem{shi2014rotational}
Y.~Shi, J.~Dobaczewski, P.~Greenlees, Rotational properties of nuclei around
  $\mathrm{No}^{254}$ investigated using a spectroscopic-quality skyrme energy
  density functional, Physical Review C 89~(3) (2014) 034309.

\bibitem{sobiczewski2007description}
A.~Sobiczewski, K.~Pomorski, Description of structure and properties of
  superheavy nuclei, Progress in Particle and Nuclear Physics 58~(1) (2007)
  292--349.

\bibitem{parkhomenko2006single}
A.~Parkhomenko, A.~Sobiczewski, Single-particle effects in decay chains of
  odd-a superheavy nuclei, International Journal of Modern Physics E 15~(02)
  (2006) 457--463.

\bibitem{CWIOK1994356}
S.~Ćwiok, S.~Hofmann, W.~Nazarewicz, Shell structure of the heaviest elements,
  Nuclear Physics A 573~(3) (1994) 356--394.

\bibitem{RevModPhys.49.833}
R.~R. Chasman, I.~Ahmad, A.~Friedman, J.~Erskine, Survey of single-particle
  states in the mass region {A} $>$ 228, Reviews of Modern Physics 49~(4)
  (1977) 833.

\bibitem{PhysRevC.110.054310}
K.~Kessaci, B.~J.~P. Gall, O.~Dorvaux, M.~Forge, A.~Lopez-Martens, R.~Chakma,
  K.~Hauschild, M.~L. Chelnokov, V.~I. Chepigin, A.~V. Isaev, et~al., Cascade
  of high-{K} isomers in ${_{102}^{255}\mathrm{No}}_{153}$, Physical Review C
  110 (2024) 054310.

\bibitem{N153}
A.~Parkhomenko, A.~Sobiczewski, Neutron one-quasiparticle states of heaviest
  nuclei, Acta Physica Polonica B 36~(10) (2005) 3115.

\end{thebibliography}


\end{document}